\documentclass[aps,prd,showpacs,superscriptaddress,groupedaddress]{revtex4}  % twocolumn submission
\usepackage{dcolumn}   % needed for some tables
\usepackage{bm}        % for math
\usepackage{microtype}
\usepackage{aas_macros}
\usepackage{times}
\usepackage{url}
\usepackage{amsmath}
\usepackage{amsbsy}
\usepackage{graphicx}
\usepackage{subfig}
\usepackage{xspace}
\usepackage{float}
\usepackage{caption}
\usepackage{amsfonts}
\usepackage{amssymb}
\usepackage{multirow}
\usepackage{color}	
\usepackage[breaklinks, colorlinks, citecolor=blue, linkcolor=black, urlcolor=black]{hyperref}%with colors
\newcommand{\ie}{\textit{i.e.,}\xspace}
\newcommand{\eg}{\textit{e.g.,}\xspace}
\newcommand{\equ}[1]{\begin{equation}#1\end{equation}}
\newcommand{\eqn}[1]{\begin{eqnarray}#1\end{eqnarray}}

\newcommand{\equref}[1]{{\xspace}Equation~(\ref{#1})}
\newcommand{\equrefs}[2]{{\xspace}Equations~(\ref{#1}) and~(\ref{#2})}
\newcommand{\figref}[1]{{\xspace}Figure~\ref{#1}}
\newcommand{\secref}[1]{{\xspace}Section~\ref{#1}}
\newcommand{\appref}[1]{{\xspace}Appendix~\ref{#1}}

\newcommand{\GP}{ \mathcal{GP}}

\newcommand{\photoz}{photo-$z$\xspace}

\renewcommand{\d}{{\mathrm{d}}}

\newcommand{\mat}[1]{\mathbf{\MakeUppercase{#1}}}

\renewcommand{\d}{\mathrm{d}}

\begin{document}

\title{\hspace*{-2mm}Data-driven, interpretable photometric redshifts trained on heterogeneous and unrepresentative data }

\author{Boris~Leistedt}
  \email{boris.leistedt@nyu.edu}
  \affiliation{Center for Cosmology and Particle Physics, Department of Physics, New York University, New York, NY 10003, USA}
  \affiliation{NASA Einstein Fellow}
  
\author{David~W.~Hogg}
  \email{david.hogg@nyu.edu}
  \affiliation{Center for Cosmology and Particle Physics, Department of Physics, New York University, New York, NY 10003, USA}
  \affiliation{Center for Data Science, New York University, 60 Fifth Avenue, New York, NY 10011, USA}
  \affiliation{Flatiron Institute, 162 Fifth Avenue, New York, NY 10010, USA}
  
\begin{abstract}
We present a new method for inferring photometric redshifts in deep galaxy and quasar surveys, based on a data driven model of latent spectral energy distributions (SEDs) and a physical model of photometric fluxes as a function of redshift. This conceptually novel approach combines the advantages of both machine-learning and template-fitting methods by building template SEDs directly from the training data. This is made computationally tractable with Gaussian Processes operating in flux--redshift space, encoding the physics of redshift and the projection of galaxy SEDs onto photometric band passes. This method alleviates the need of acquiring representative training data or constructing detailed galaxy SED models; it requires only that the photometric band passes and calibrations be known or have parameterized unknowns. The training data can consist of a combination of spectroscopic and deep many-band photometric data, which do not need to entirely spatially overlap with the target survey of interest or even involve the same photometric bands. We showcase the method on the $i$-magnitude-selected, spectroscopically-confirmed galaxies in the COSMOS field. The model is trained on the deepest bands (from SUBARU and HST) and photometric redshifts are derived using the shallower SDSS optical bands only. We demonstrate that we obtain accurate redshift point estimates and probability distributions despite the training and target sets having very different redshift distributions, noise properties, and even photometric bands. Our model can also be used to predict missing photometric fluxes, or to simulate populations of galaxies with realistic fluxes and redshifts, for example. This method opens a new era in which photometric redshifts for large photometric surveys are derived using a flexible yet physical model of the data trained on all available surveys (spectroscopic and photometric).
\end{abstract}


\maketitle

%%%%%%%%%%%%%%%%%%%%%%%%%%%%%%%%%%%%%%
\section{Introduction}

Ongoing and upcoming photometric galaxy surveys such as the Dark Energy Survey \citep[DES,][]{Abbott:2005bi}, The Kilo-Degree Survey \citep[KIDS,][]{deJong:2013}, and LSST \citep{Abell:2009aa} will allow us to probe the content, dynamics, origins, and fate of the universe at unprecedented accuracy \citep[see \eg][]{Peacock:2006kj, Weinberg:2012es}. They will prove essential to uncover the properties of dark matter, measure the properties of high-energy particles such as neutrinos, and test models of the early universe, gravity, and the late-time accelerated expansion.
But exploiting those surveys for cosmological tests requires estimating the redshift of each extragalactic object from a handful of noisy photometric flux measurements.
This challenging task is commonly known as photometric redshift (\photoz) estimation.
Photo-$z$ estimates are typically used to group galaxies into bins of redshift and other informative properties such as color or morphology.
It is by measuring the statistical properties of those groups and confronting them with theoretical predictions that cosmological models can be tested.
Typical observables of interest include the auto-and cross-correlation of galaxy positions and shapes as a function of sky separation, redshift, color, or luminosity.

Photometric redshifts are currently the dominant source of both statistical and systematic errors in the analysis of photometric galaxy surveys.
This is because the number of objects and the physical volume probed by those surveys have steadily grown in the past decades, and their exploitation is now limited by the systematic errors made while processing and exploiting the data.
For instance, those can result from the complicated reduction of raw images, the imperfect discrimination between various types of objects (\eg stars, galaxies, quasars), the incomplete understanding of the survey selection function, or the inaccurate estimation of statistics such as 2-pt correlation functions.
Yet, imperfect photometric redshift estimation significantly exceeds all other sources of errors.
This is fueled by the unavailability of validation data sets or realistic survey simulations for validating \photoz algorithms at the precision needed.
Those issues raise difficult challenges for the exploitation of ongoing experiments and may jeopardize future surveys if they are not overcome.
We now briefly outline the advantages and limitations of the two main classes of algorithms used for estimating redshifts from photometric fluxes: template fitting and machine learning methods.
Note that in what follows we will focus on galaxies, but it should be clear that the discussions and methods below are equally applicable to other extragalactic objects such as quasars.

In machine learning methods \citep[\eg][]{Kind:2013eka, Collister:2003cz, Sadeh:2015lsa}, the relation between galaxy fluxes and redshifts is fitted using a very flexible model, the details of which depend on the algorithm under consideration (\eg neural networks, random forests, etc).
This model must be trained on ancillary data where true redshifts are available; those are typically high-quality estimates from a spectroscopic survey.
This approach is very powerful for learning complicated relationships in the data without designing and validating a potentially complex physical model.
For this reason, it usually provides excellent redshift estimates in the regions with good training data, even in the presence of imperfect fluxes (\ie biased or with underestimated errors).
In other words, in the interpolation regime, machine learning methods excel.
However, they require the training data to be representative of the target data, \ie to have identical redshift and flux distributions.
As a result, they perform poorly in regions with few or no training data.
This is a significant issue since training sets (\ie galaxy surveys with spectroscopic redshifts) are much shallower than the target photometric surveys.
This can be alleviated via re-weighting, but still requires the training to be similar to the target data, and to be based on the same photometric band-passes.
Generally speaking, machine learning methods do not know about the underlying physics of the problem: flux measurements arise from observing a reshifted galaxy spectrum through known photometric band-passes.
They will partially learn those effects from the training data, but they are not required to, which limits their robustness in regimes critical to cosmological applications.

Template-fitting methods \citep[\eg][]{Benitez:1998br, Brammer:2008qv, Feldmann:2006wg}  address this issue. 
If a library of galaxy spectra (\ie templates for the spectral energy distributions of various galaxy types) is available, then one can solve for the redshift and type of a galaxy given the observed photometric fluxes.
A significant advantage over machine learning methods is the ability to perform the fit in a fully probabilistic fashion, with explicit priors over the types and redshifts of galaxies.
While the outputs of some machine learning methods can be interpreted in probabilistic terms, most of them implicitly construct complicated priors from the training data (related to the representativeness issue mentioned above), making the probabilistic interpretation difficult.
Template fitting approaches provide an elegant solution to estimate photometric redshifts and also other galaxy properties (\eg star formation history).
However, they are very restrictive; one has to assume that all galaxies observed in the survey of interest can be described by the library of templates or the physical model used. 
The complexity and imperfections of observed fluxes (\eg biases or underestimated errors) cannot easily be captured either.
Methods have been developed to relax these issues (\eg by introducing correction terms to existing template libraries or adopting very flexible spectral template with numerous physical parameters), but are expensive and do not bring template fitting to the accuracy level needed by modern photometric surveys.

A third class of methods for estimating photometric redshifts exists, and is sometimes referred to as ``clustering redshifts" \citep[\eg][]{Matthews:2010an}.
It exploits spatial information and the proximity of galaxies in real space (sky position and redshift) to constrain the redshifts of galaxies.
Since it does not directly uses flux information we will not discuss it further, and we will assume that any flux-based \photoz method (such as the one presented below) could be improved by adding spatial information.

Recent analyses of galaxy surveys (DES, KIDS) and various detailed \photoz investigations have highlighted the advantages and drawbacks discussed above \citep[see \eg][]{Newman:2013cac, Dahlen:2013fea, Sanchez:2014zgq, Schmidt:2014ela, Bonnett:2015pww}. 
%Typically, machine learning methods \citep[\eg][]{Kind:2013eka, Collister:2003cz, Sadeh:2015lsa} provide accurate redshift estimates for bright, low redshift galaxies, but are unreliable at high redshift or faint magnitudes because of the lack of training data.
%Template fitting methods \citep[\eg][]{Benitez:1998br, Brammer:2008qv, Feldmann:2006wg} behave well at high redshift since the underlying physical model allows to extrapolate.
%However, their redshift estimates suffer from various systematics due to the limitations of the template sets in existence, which are small and constructed from low redshift galaxies.
%They are also very sensitive to the prior knowledge about the abundance and redshift dependence of the various types modeled.
Most of those could be (and are being) resolved via time-consuming validation and ad-hoc corrections, but they demonstrate the limitation of standard \photoz techniques.
The method presented here is an attempt at addressing those issues altogether and harnessing the flexibility of both machine learning and template fitting, via the construction of a large collection of latent SED templates from the training data.
We will present a generic approach to perform this construction directly in flux--redshift space without explicitly going through galaxy spectra, by exploiting Gaussian Processes with kernels capturing the effect of cosmological redshift and the projection onto photometric band-passes.
In other words, we are efficiently constructing a large set of physical flux-color-redshift models.
This is similar to the K-correct model \citep{Blanton:2007} but incorporating model uncertainties and derived from a much larger data set consisting of photometric data with spectroscopic redshifts.
As demonstrated below, this model is straightforward to train and validate, and it does not require the training data to have the same photometric bands or redshift and noise distributions as the target data.
A central conclusion of this paper is to show that template fitting can be improved by directly constructing templates from heterogeneous training data with weak or no assumptions about galaxy SEDs.

The remainder of this paper is structured as follows: In \secref{sec:methods}, we present our novel \photoz inference method and we discuss its advantages and limitations. We illustrate its performances on real data in \secref{sec:data}, and we conclude in \secref{sec:concl}. The Appendices of this paper provide useful technical details about our implementation. $\mathcal{N}(a;A)$ denotes the Gaussian distribution of mean $a$ and covariance $A$ (we will use this convention for both univariate and multivariate distributions).

%%%%%%%%%%%%%%%%%%%%%%%%%%%%%%%%%%%%%%
\section{Photometric redshift inference via physical Gaussian Processes}\label{sec:methods}

%%%%%%%%%%%%%%%%%%%%%%%%%%%%%%%%%%%%%%%%%
\subsection{Background and assumptions}

For the purpose of this work, a galaxy is fully described by its rest-frame luminosity density or spectral energy distribution (SED), denoted by $L_\nu(\lambda_\mathrm{em})$ at the emitted wavelength $\lambda_\mathrm{em}$.
For this same galaxy at redshift $z$, the flux observed today at a wavelength $\lambda_\mathrm{obs}$ reads
\equ{
	f_\nu(\lambda_\mathrm{obs}, z) = \frac{(1+z)}{4\pi D^2(z)}  \ L_\nu\left(\frac{\lambda_\mathrm{obs}}{1+z}\right)
}
where $D(z)$ is the luminosity distance \cite{Hogg:2002yh}. 
%Three effects are in play: the Doppler shift of photons, the dimming of the flux due to distance, and its amplification due to the background expansion of the universe. 
We are interested in photometric measurements of the flux $f_\nu$ in a set of bands $b=1, \dots, N_b$ described by a set of filter responses $\{ W_b(\lambda) \}$, which we assume to be known.
The photometric flux measured in the $b$-th band is
\eqn{
	F_b(z) &=& \frac{1}{g^\mathrm{AB} C_b}   \int_0^\infty f_\nu(\lambda_\mathrm{obs}, z) \ W_b(\lambda_\mathrm{obs}) \frac{\d\lambda_\mathrm{obs}}{\lambda_\mathrm{obs}}  \\
		&=& \frac{(1+z)^2}{4\pi D^2(z) g^\mathrm{AB} C_b}   \int_0^\infty L_\nu(\lambda_\mathrm{em}, z) \ V_b\bigl(\lambda_\mathrm{em}(1+z)\bigr) \ \d\lambda_\mathrm{em} \label{fluxredshift}
}
where $g^{AB}$ is the zero point of the AB photometric system, and $C_b$ is the filter normalization constant $C_b = \int_0^\infty W_b(\lambda) \d\lambda / \lambda$. The change of convention from $W_b$ to  $V_b(\lambda) = W_b(\lambda)/\lambda$ will simplify some of the calculations below.

We are interested in estimating the redshifts of a set of {\bf target galaxies}, for which we have noisy photometric flux measurements $\hat{\mat{F}}=(\hat{F}_1, \dots, \hat{F}_{N_b})$ with known variance (\eg Gaussian errors). 
We assume that a {\bf training set} is available, \ie an other set of galaxies with noisy photometric flux measurements.
Furthermore, we assume that the redshifts of those training galaxies are available.
Those are typically obtained via high-resolution measurement of $f_\nu$ and subsequent estimation of the type and redshift of the object.
Those high-quality estimates are often called spectroscopic redshifts.
In what follows they are assimilated to the true redshifts, although the method could (and will) be extended to support redshift errors, as discussed in \secref{sec:discussion}. 

We also assume that a set of library of galaxy SED templates is available. 
Standard template fitting methods would directly rely on it to estimate redshifts for the target galaxies. 
Typical machine learning methods wouldn't exploit it at all. 
The method presented here weakly relies on the template SED library to guide the learning of the model. 
This point will be extensively discussed in \secref{sec:discussion}, where we will also detail the other assumptions of the method.

%%%%%%%%%%%%%%%%%%%%%%%%%%%%%%%%%%%%%%%%%
\subsection{Inference}

We introduce a variable $t$ labelling galaxy types, described by a (continuous or discrete) ensemble of SEDs $L_\nu(\lambda, t)$, so that the photometric fluxes become $F_b(z, t)$. 
For a target galaxy of interest, the posterior distribution on its redshift given noisy photometric flux measurements $\hat{\mat{F}}$ is
\eqn{
	p(z | \hat{\mat{F}}) \propto \int \mathrm{d}t\  p\bigl(\hat{\mat{F}}| z, t\bigr)\ p(z, t)
  \approx \sum_{i}   p\bigl( \hat{\mat{F}}| z, t_i\bigr) \ p(z | t_i) \ p(t_i) \label{eq:redshiftposterior}
  }
This is identical to the approach adopted in standard template fitting methods, where $t_i$ with $i=1, \cdots, N_T$ labels the SED templates, and $p(z | t_i) p(t_i)$ capture prior information about their redshift distributions and abundances (often calibrated on training data).
  
We take a different approach and construct a set of templates from the training set itself. 
The type $t_i$ is constructed from the $i$-th training galaxy, \ie with its noisy photometric fluxes $\hat{\mat{F}}_i$ and the spectroscopic redshift $z_i$. 

Hence, for each pair of target and training galaxies, we are going to be able to write
 \eqn{ 
 	p\bigl(\hspace*{-2pt}\underbrace{\hat{\mat{F}}| z}_{\mathrm{target}}, t_i \bigr) &=& p\bigl(\hat{\mat{F}}| z, \underbrace{z_i, \hat{\mat{F}}_i}_{\mathrm{training}} \bigr) \\
	&=& \int\d \mat{F} \ p\bigl(\hat{\mat{F}}| \mat{F}\bigr) \ p\bigl( \mat{F}| z, z_i, \hat{\mat{F}}_i\bigr).\label{eq:traintarpair}
	}
The first term is the flux likelihood function, comparing the noisy fluxes of the target galaxy with the (noiseless) model fluxes computed from the training galaxy. 
The second term is a prediction for the fluxes of the training galaxy {\it at a different redshift} $z$ (not $z_i$!). 
It will be discussed in the next section; we first discuss the likelihood function.

The simplest likelihood function we could write has uncorrelated Gaussian flux errors on the observations, and becomes a simple product of univariate Gaussian distributions.
However, this likelihood function is too simplistic.
First, it ignores model uncertainties, which will arise from the model fluxes $\mat{F}$ we will construct in the next section from the second term of \equref{eq:traintarpair}.
As we will see, the output of our predictions, based on Gaussian Processes, will be multivariate Gaussian, \ie $p\bigl( \mat{F}| z, z_i, \hat{\mat{F}}_i\bigr) = \mathcal{N}(\mat{F}- \mat{F}^*;\mat{\Sigma}^*_\mat{F})$ with $\mat{F}^*$ and $\mat{\Sigma}^*_\mat{F}$ the mean predictions and their variance, which are functions of $z$, $z_i$ and $\hat{\mat{F}}_i$, detailed in the next section.
Second, we must introduce a factor $\ell$ scaling the model and its uncertainties. 
This is to account for the potential difference in absolute luminosity of galaxies of the same type (\ie different normalization of the same SED).
One extreme way to implement this step would be to proceed with a likelihood relying on colors, \ie ratio of fluxes.
However, the absolute luminosities of the target and the training galaxies carry a significant amount of information, which we will exploit in the pairwise comparison performed in \equref{eq:traintarpair}.
We assume that the errors on the observations are Gaussian and described by a covariance matrix $\mat{\Sigma}_{\hat{\mat{F}}}$. 
With these assumptions, we derive the likelihood function 
\eqn{
	p\bigl(\hat{\mat{F}}| z, z_i, \hat{\mat{F}}_i \bigr)  &=& \int \d\ell \ \mathcal{N}\left( \hat{\mat{F}} - \ell\  \mat{F}^*; \mat{\Sigma}_{\hat{\mat{F}}} + \ell^2 \mat{\Sigma^*_{{\mat{F}}}}  \right)\ p(\ell ). \label{eq:likelihood}
}
In our experiments we found that adopting a the prior $p(\ell ) =  \mathcal{N}(1; \sigma^2_\ell)$ is a good approach since it accounts for the proximity of pairs of galaxies in absolute luminosity.
The optimization of hyperparameters such as $\sigma_\ell$ will be discussed below. 
Further details about our implementation of \equref{eq:likelihood} are provided in \appref{app:fluxlikelihood}.

%%%%%%%%%%%%%%%%%%%%%%%%%%%%%%%%%%%%%%%%%
\subsection{Gaussian Process in flux--redshift space}\label{sec:gppres}

Let us consider the second term of \equref{eq:traintarpair}, which requires to use the noisy fluxes of the training galaxy $\hat{\mat{F}}_i$ at redshift $z_i$ and predict (noiseless, model) fluxes at a different redshift $z$.
In other words, we must compute the probability that the target galaxy has the same SED as the training galaxy but at a different redshift. 
An elegant way to address this problem could be to explore the set of SEDs compatible with the photometry $\hat{\mat{F}}_i$ at redshift $z_i$, then redshift all of those individually at redshift $z$. Finally, we could compute the mean and variance of the predicted fluxes, and compare them to the noisy fluxes of the training galaxy.
The SED model could be arbitrarily complex, \eg derived from a model of galaxy formation, synthetic templates, etc. 
However, this approach is computationally intractable since it requires to simulate large numbers of SEDs and integrating those for comparison to each training galaxy and for predicting fluxes at several other redshifts (\eg via MCMC sampling methods). 
In addition, given the broadness of photometric bands and the typical flux errors, the predicted fluxes are likely to be relatively insensitive to the details of the SED model. 
In other words, a complicated SED model is probably unnecessary for the purpose of estimating redshifts for wide-area broad-band surveys such as DES and LSST.
To resolve these issues, we will use a Gaussian Process $F(b, z) \sim \mathcal{GP}\bigl( \mu^F, \ k^F\bigr)$, and encode the SED model, its redshifting, and its projection onto the photometric band-passes in the mean function $\mu^F$ and the kernel $k^F$. 

Gaussian Processes are a flexible method for fitting noisy data and making predictions in both the interpolation and extrapolation regimes (\ie where there is and isn't training data).
In fact, they encompass a range of widespread methods, from simple linear models to neural networks, and produce well-defined model predictions and uncertainties.
When the likelihood function is Gaussian (\eg with Gaussian noise), most operations on Gaussian Processes, including posterior distribution and marginal likelihood calculations, are analytically tractable.
This makes them extremely appealing.
In particular, the posterior distribution is a multivariate Gaussian, which is adequate for the flux likelihood function of \equref{eq:likelihood}. 
One of their main drawback is the potentially large matrix operations to predict the mean and the covariance of the outputs (here $\mat{F}^*$ and $\mat{\Sigma^*_{{\mat{F}}}}$). 
However, this will not be a problem here since we fit each training galaxy with a separate Gaussian Process.
Thus, the number of data points to fit is merely the number of photometric bands, which is small (from a few to tens).
For further details about Gaussian Processes, we direct the reader to the excellent introduction of Ref.~\cite{Rasmussen:2005}.

A significant advantage of Gaussian Process is the ability to specify a mean function and a covariance kernel that capture the known or expected correlations of the problem at hand, for both the signal and the noise.
While classical mean functions and kernels could be used here, they would predict fluxes $F_b(z, t_i)$ that wouldn't correspond to an SED being redshifted and projected onto the band-pass $W_b$.
Instead, we will impose the mean function $\mu^F$ and the kernel $k^F$ to capture the expected correlations across redshift and bands resulting from the know setup and physics of the problem: the fluxes result from observing SEDs through filter responses $\{ W_b(\lambda)\}$, and those SEDs are redshifted according to \equref{fluxredshift}.  
Concretely, we want to define a mean function and a kernel that implicitly solve the same procedure described above: constructing SEDs compatible with the $\hat{\mat{F}}_i$, redshifting and integrating them to obtain flux predictions $\mat{F}$. 
It is possible under certain assumptions and descriptions of the SEDs, as described below.

We model the latent, underlying SED of each training galaxy as a linear mixture of templates $T^t_\nu(\lambda)$  with $t=1, \cdots, N_T$ (taken from the existing template library) and residuals that take the form of a zero-mean Gaussian Process $R_\nu \sim \mathcal{GP}\bigl(0, k^\lambda(\lambda,\lambda') \bigr)$. In other words,
\eqn{
	L_\nu(\lambda, \bm{\alpha}, \ell) \ = \  \underbrace{\ell \ \sum_{t=1}^{N_T} \alpha_t\ T_\nu^t(\lambda)}_{\rm templates} + \ \underbrace{ \ell \ R_\nu(\lambda)}_{\rm residuals} \ \ \sim \ \ \mathcal{GP}\Bigl(\ell \ \sum_t \alpha_t T^t_\nu(\lambda), \ \ell \ell' \ k(\lambda, \lambda')\Bigr), \label{eq:sedmodel},
}
with $\bm{\alpha}=(\alpha_1, \dots, \alpha_{N_T})$ the template coefficients. 
$\ell$ is the absolute luminosity and allows us to scale the residuals for each galaxy. 
It is essential for making sure the residuals are correctly normalized and have roughly the same amplitude for all training galaxies. 
Without the extra factor $\ell$, the prior on the Gaussian Process residuals (\eg its variance) would affect low-luminosity galaxies more than high-luminosity ones.
Note that $\ell$ could be fixed or optimized in various ways; we discuss one in the data application of \secref{sec:data}.

Inserting the model of \equref{eq:sedmodel} in \equref{fluxredshift}, we find that the photometric fluxes are also a Gaussian Process,
\equ{
	F_b(z, \bm{\alpha}, \ell) \sim \ \mathcal{GP}\Bigl( \mu^F(b, z, \bm{\alpha}), \ k^F(b,b',z,z',\ell,\ell')\Bigr). \label{eq:fluxgp}
	}
It is straightforward to show that the mean function is
\eqn{
	\mu^F(b, z, \bm{\alpha}) &=& \frac{\ell (1+z)^2 }{4\pi D^2(z)g^\mathrm{AB} C_b}  \sum_{t=1}^{N_T} \alpha_t \int_0^\infty T_\nu^t(\lambda) \ V_b\Bigl((1+z)\lambda\Bigr) \ {\d\lambda} 
	\ =  \   \ell \sum_{t=1}^{N_T}  \alpha_t F^t_b(z) . \label{eq:gpmeanfct}
}
 $F^t_b(z)$ denotes the $b$-th flux of the $t$-th template at redshift $z$, which can be pre-computed once the initial library of templates is chosen. 
In fact, this is the cornerstone of standard template fitting methods, which would simply use $F^t_b(z)$ in the flux likelihood function.

The covariance function or kernel of the Gaussian Process is
\eqn{
	k^F(b,b',z,z',\ell,\ell') &=& \left( \frac{ (1+z)(1+z') }{4\pi D(z) D(z') g^\mathrm{AB}} \right)^2  \frac{\ell\ell'}{C_bC_{b'}} \int_0^\infty \ V_b\Bigl((1+z)\lambda\Bigr) \ V_{b'}\Bigl((1+z')\lambda'\Bigr) \ k(\lambda, \lambda') \ \d\lambda\d\lambda' . \label{eq:gpgenkernel} 
}
This is the general form for a kernel acting on fluxes as a function of band and redshift, corresponding to the conversion of a kernel $k(\lambda, \lambda')$ living in SED space (see Ref.~\cite{Miller:2015:gpqso} for a specific example, applied to quasar spectra and photometry).
While the mean function of \equref{eq:gpmeanfct} is trivial since $F^t_b(z)$ is pre-computed, the kernel of \equref{eq:gpgenkernel} must be numerically evaluated since the photometric filter responses are usually experimentally measured and tabulated. 
This calculation is challenging because it must be performed for all pairs of fluxes and all training galaxies. 
But for some specific choices of $k$ and representations of $V_b$, closed analytical forms of $k^F$ will exist. 
We give such form in \appref{sec:rbfgp}, which we use in the remainder of this paper. 
It is based on a radial basis function (RBF) kernel $k\propto \exp(-\beta(\lambda-\lambda')^2)$ and approximations of the filter responses with Gaussian mixtures.
In this case, one can exactly compute $k^F$ for any inputs without resorting to numerical integration methods.

Now that we derived a suitable Gaussian Process, we can write $p\bigl( \mat{F}(z,t_i) | z_i, \hat{\mat{F}}_i\bigr)$ as the flux predictions when fitting the $i$-th training set galaxy. 
It is a multivariate Gaussian, a standard Gaussian Process prediction, \eg following the techniques described in Ref.~\cite{Rasmussen:2005}. 
For completeness we provide its full form as well as further technical details in \appref{sec:gppred}.

%%%%%%%%%%%%%%%%%%%%%%%%%%%%%%%%%%%%%%%%%
\subsection{Discussion}\label{sec:discussion}

Our method can be summarized as follows.
The posterior distribution on the redshift of a target galaxy is obtained via a pairwise comparison with training galaxies, following \equref{eq:redshiftposterior}. 
For each training-target pair, we evaluate \equref{eq:traintarpair}, the probability that the training and the target galaxies actually have the same SED, but they may simply be at different redshifts.
The predictions for the fluxes of the training galaxy at $z$ are calculated via a Gaussian Process. 
Finally, the likelihood function compares the noisy fluxes of the target with the model fluxes of the training galaxy, following \equref{eq:likelihood}.

The method is flexible in multiple ways: the Gaussian Process has several parameters (see \appref{sec:gppred}), so do the luminosity and redshift priors appearing in \equrefs{eq:redshiftposterior}{eq:likelihood}.
Those can all be optimized on the training data to improve the quality of the resulting redshift estimates according to various metrics of interest (\eg mean redshift, confidence intervals, etc).
This will be presented in detail in \secref{sec:data} in the context of a specific data set.

\smallskip
We now discuss the assumptions and limitations of our method.
\smallskip

\paragraph{Flux biases and band passes} 
We have assumed that the flux measurements are unbiased, and that their errors are Gaussian and correctly characterized.
However, real flux measurements and their error estimates exhibit magnitude-, type- and redshift-dependent biases.
Furthermore, photometric filter curves may be mischaracterized, which introduces extra biases.
Since this method is trained on real fluxes, it will be able to absorb some of these biases. 
In addition, it is straightforward to add (hyper)parameters describing such biases.
One can then fit or marginalize for those while training the method.
For example, in the demonstration of the next section we have included a parameter describing an extra flux error term added in quadrature to all flux errors.
Biases in the characterization of the photometric band passes could also be parametrized and inferred from the training data, if necessary.

\paragraph{Negative flux predictions}
In the description above, the latent SED and the fluxes are not formally constrained to be positive. 
However, this is not really a problem since fluxes will go only negative in regimes where the SED is not well constrained.
The large errors will prevent those negative fluxes from having any effect on the results, since their contribution to the likelihood function is negligible.

\paragraph{Availability of templates} 
The use of a template library in the SED model of \equref{eq:sedmodel} (or equivalently, using a non-zero mean function for the Gaussian Process) is not at all necessary.
Gaussian Processes could well be used to model SEDs without mean functions, which we have checked in various experiments.
This is because the posterior distributions obtained using a zero-mean Gaussian Process is non-zero.
However, we find that including the template library in the mean function helps guiding the Gaussian Process and minimizing the amount of residuals needed to fit each training galaxy.

\paragraph{Simple SED model}
Our SED model is simple and assumes that the SEDs of all galaxies in the training and target data can be described by a linear mixture of templates with smooth corrections.
The kernel for these corrections only includes smooth continuum corrections as well as lines at specific locations.
While this is very restrictive, it encompasses a large space of galaxy SEDs and is sufficient for fitting broad-band photometric observations; we find that opting for smaller correlation lengths and more lines does not affect the quality of the results shown below.
However, the model can easily be extended to more complicated forms: the mean function could include a larger library of templates, and the kernels could be improved in various ways \cite[see \eg][]{Miller:2015:gpqso}.
Both could be directly learned from real galaxy spectra, for example.

\paragraph{Availability of redshifts} 
At present our method requires photometric fluxes and spectroscopic redshifts for all training galaxies. 
In future work it will be extended to support redshift errors, for both spectroscopic and photometric redshifts in the training set. 

\paragraph{Use of galaxy spectra}
In this first paper, our approach does not exploit spectroscopic measurements of $f_\nu$ for the training galaxies, unlike methods such as K-correct \cite{Blanton:2007}.
Those could be used in place of the latent SED model inferred by the Gaussian Process. 
However, the mapping between measured photometric fluxes and spectroscopic SEDs can be complicated and not satisfy \equref{fluxredshift}, for example due to the way they are measured in data.
For this reason, we currently focus on using photometric flux measurements only.

\smallskip
We now turn to the advantages of our novel approach.
\smallskip

\paragraph{Data-driven, physical modeling of photometric fluxes}
The approach described here is similar to doing machine learning constrained by the effect of redshifts on galaxy SEDs.
Indeed, correlations between fluxes at various redshifts and bands are predictable since fluxes are deterministically connected to SEDs.
Most machine learning methods will ignore this and fit for redshifts and fluxes in the training data with extremely flexible functions.
However, not all combinations of fluxes are allowed; in fact, given observations of a training galaxy, the space of possible fluxes at other redshifts and band is drastically reduced.
Our method correctly incorporates this information because it implicitly (via the Gaussian Process) models SEDs and only produce fluxes that result from integrating those SEDs.
Assuming there are no biases in the fluxes or the redshifts of the training galaxies (see remarks above), this is a provably correct scheme for constructing a data-driven model from the training data with an implicit system of templates.
Unlike standard template fitting, which relies on a small set of SED templates constructed with low-redshift data or complex physical models, this method directly constructs a large set templates from the training data, with weak assumptions about galaxy SEDs.

\paragraph{Interpretable model and probabilities} 
As evident by our presentation of the method, it produces interpretable probability distributions regardless of the training or target galaxies under consideration, or the values of the various (hyper)parameters.
This is a powerful property: the model is flexible, yet physical.
It consists of SEDs (in fact, flux--redshift relations corresponding to SEDs) constructed from the measured fluxes and redshifts of the training galaxies.
The probabilities this model provides for the redshifts of target galaxies are governed by a few equations and easily interpretable. 
For instance, one can isolate the contribution of each training galaxy, which will be illustrated in the next section.
Furthermore, the probabilities are correctly normalised; the marginalized likelihood or evidence can be calculated by integrating \equref{eq:redshiftposterior}. 
This provides a measure of the goodness of the prediction for each target galaxy.
One could consider a different training set and run standard Bayesian model comparison tests. 

\paragraph{Flexible (hyper)parameters}
The various hyperparameters of the Gaussian Process and priors can be optimized to improve any metric of interest.
Obvious metrics include the quality of the mean or peak of the redshift posterior distribution of \equref{eq:redshiftposterior}, over the entire training set or an external validation set.
But less standard metrics could be adopted, such as the quality of the confidence intervals or the redshift distributions.

\paragraph{Speed and storage}
If one decides to expand the training set, the posterior distribution of each target galaxy can simply be updated by adding terms corresponding to the new training galaxies since \equref{eq:redshiftposterior} is linear.
Of course, one may wish to re-optimize the hyperparameters of the method if the training set significantly changes. 
Another powerful property resulting from \equref{eq:redshiftposterior} is the ability to truncate the sum and only keep the training galaxies that significantly contribute to the result.
We find that this compression can be very efficient: typically, only a few (\ie a fraction of the full set) training galaxies contribute to the posterior distribution of each target galaxy.
By remembering which ones, one can in fact quickly compute a very good approximation of the posterior distribution. 
In fact, one can also identify which training galaxies do not contribute to any targets and could be dropped from the training set without affecting the results.
Finally, all operations, from the construction of GP models to the training-target pairwise comparisons, can be performed in parallel; only \equref{eq:redshiftposterior} requires an aggregation of the results.
 
\paragraph{Heterogeneous, incomplete training sets} 
One of the most important features of this approach is the ability to exploit heterogeneous data sets for the training set.
This is because \equref{eq:traintarpair} does not require the training and target galaxies to share properties such as photometric bands or noise.
In other words, the Gaussian Process we constructed is agnostic to the actual bands and noise in both the training and the data: as long as the filter responses of all the bands considered are known, one can fit and predict any combination of interest.
Of course, for any given training galaxy, the variance of the flux predictions as a function of redshift will strongly depend on the bands and noise under consideration. 
For instance, the availability of a flux will inform the flux predictions in redder bands at higher redshift.

%%%%%%%%%%%%%%%%%%%%%%%%%%%%%%%%%%%%%%
\section{Demonstration on SDSS data}\label{sec:data}

We now showcase the features and flexibility of the novel method on data. 

%%%%%%%%%%%%%%%%%%%%%%%%%%%%%%%%%%%%%%
\subsection{Setup}

We consider the G10/COSMOS data \citep{Davies:2015}, publicly available at \url{http://cutout.icrar.org/G10/dataRelease.php}. 
It is a recent compilation of data covering the COSMOS field, which involves a rich set of deep photometric observations as well as a spectroscopically confirmed objects.
We do not consider the reprocessed version of Ref.~\citep{Andrews:2016}; instead we consider the version of Ref.~\citep{Davies:2015}, where flux measurements from various surveys are collated without further adjustments or reprocessing.
We make this choice to demonstrate that we can deal with the challenging systematics arising from combining fluxes from various sources, thus with heterogeneous calibration and noise properties.
Few surveys have the opportunity to fully reprocess existing data to produce consistent flux measurements.  
We consider the subset of objects with good spectroscopic redshifts, which are mostly from the SDSS, VVDS, and PRIMUS surveys, as detailed in Ref.~\citep{Davies:2015}.
We select 9,923 training and 8,699 target galaxies, according to the criteria below.

\begin{figure}
\centering\begin{tabular}{cc}
\begin{minipage}{8cm}\includegraphics[width=8cm]{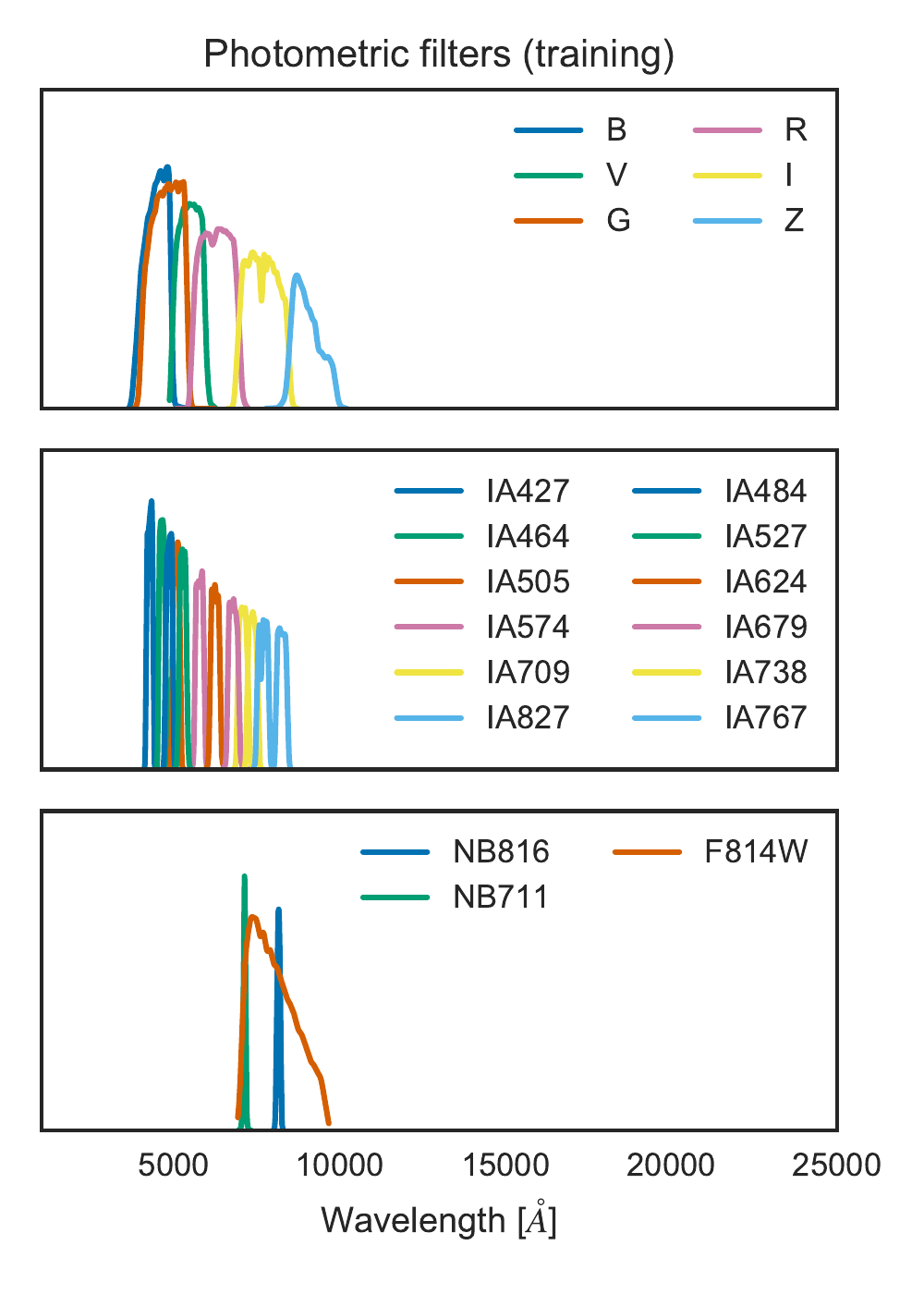}\end{minipage}&
\begin{minipage}{8cm}\includegraphics[width=8cm]{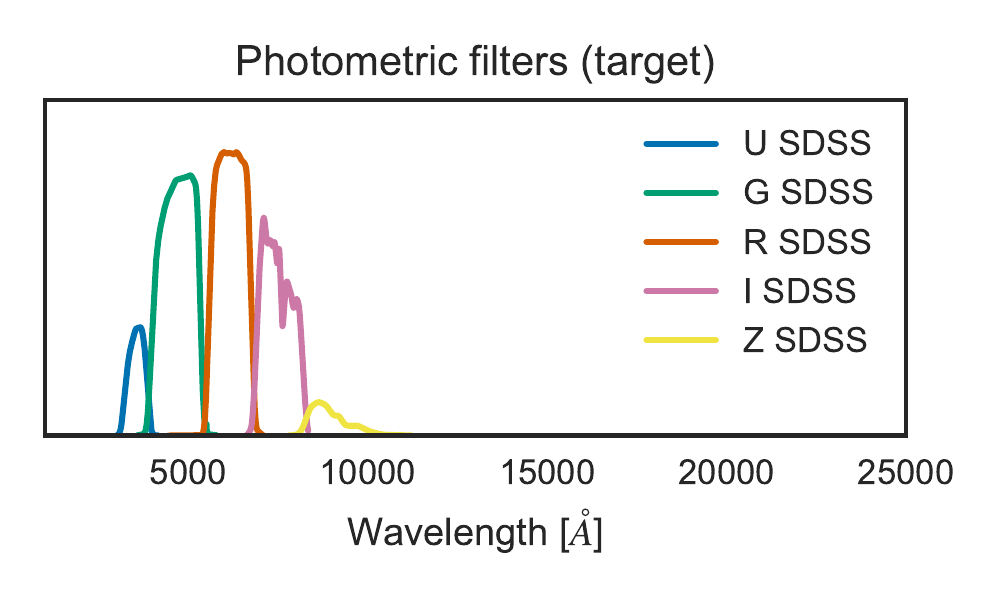}\end{minipage}
\end{tabular}
\caption{Photometric filters used for the data demonstration.}
\label{fig:filters}
\end{figure}

For the training set, we consider the following bands, shown in \figref{fig:filters}: B V G R I Z (SUBARU broad bands), NB816 IA427 IA464 IA505 IA574 IA709 IA827 NB711 IA484 IA527 IA624 IA679 IA738 IA767  (SUBARU intermediate and narrow bands), and F814W (HST). 
We require all fluxes to have a signal-to-noise greater than 2. 
For the validation/target set, we only take the five $ugriz$ SDSS broad bands.
We require the $r$ and $i$ fluxes to have a signal-to-noise greater than 5.
We apply extinction corrections to all fluxes, according to the $E(B-V)$ coefficients from Ref.~\citep{Schlegel:1998} and the conversion factors shown in Ref.~\citep{Andrews:2016}.
The objects passing the criteria of both the training and target sets are put in the target set, which leaves us with a balanced 9,923 objects for training and 8,699 for validation.

\begin{figure}
\centering
\includegraphics[width=16cm]{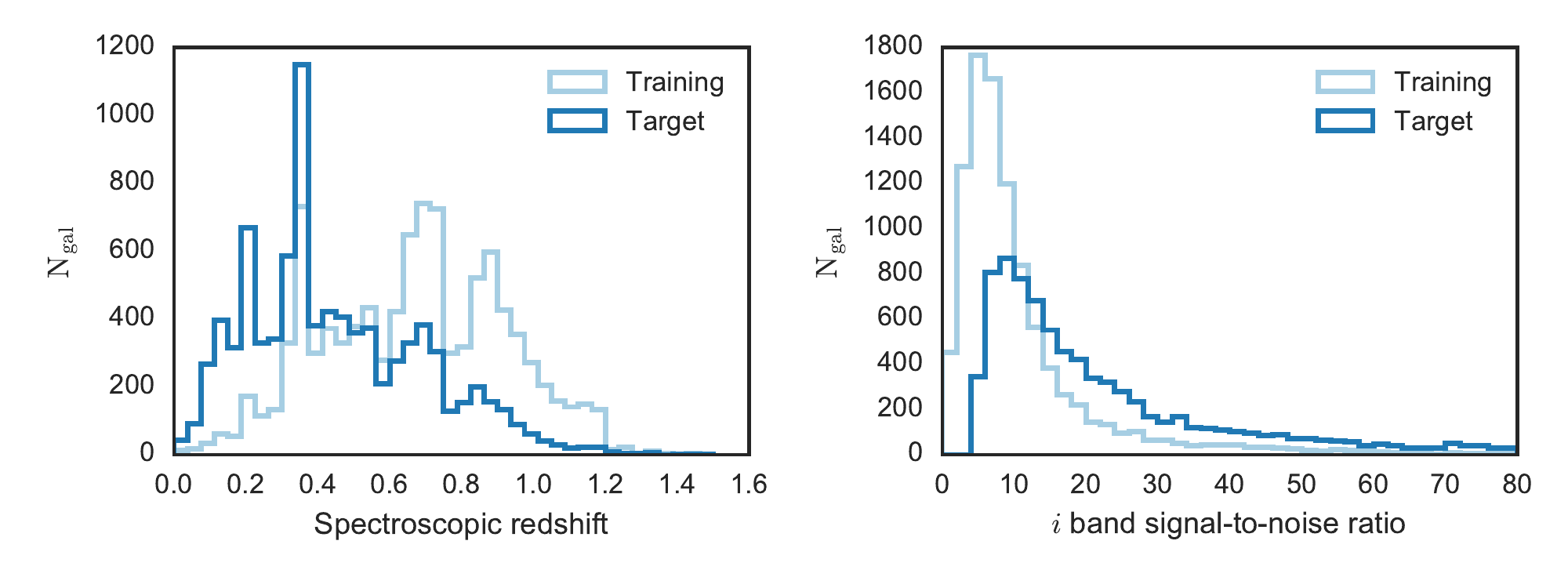}
\caption{Distributions of the redshifts and SDSS $i$ band signal-to-noise ratios for the galaxies in the training and target sets considered. The training set is deeper than the target set, and its $i$ band magnitude is noisier. Luckily, deeper SUBARU+HST observations are available, and will be exploited to build template SEDs.}
\label{fig:training_vs_target}
\end{figure}

The properties of the training and target data are drastically different.
Not only the filter sets are different, as shown in \figref{fig:filters}, but the redshift distribution of the training data is deeper and more structured than the target.
This is due to the cuts described above, as demonstrated in \figref{fig:training_vs_target} which also shows that the signal-to-noise in the $i$ band .
The training set is significantly deeper and more diverse and heterogeneous than the target, due to the complicated cuts and selection effects on 21 deep bands compared to the noisier 5 SDSS bands.
Yet, we should be able to use the deeper training data to construct a flexible model and infer photometric redshifts for the target.
Note that we also make use of the SDSS bands in the training set for cross-validation and optimization purposes, as described below.

We will run our new method and compare it to standard template fitting, to illustrate that a much richer space of templates and more flexible priors significantly improves the quality of the \photoz estimates.
In both cases, we use the 8 classic SED templates from the CWW library \citep{Coleman:1980ej, Benitez:1998br}. 
They are the central ingredient to standard template fitting, while in our method they form the mean function of the Gaussian Process and weakly affect the results. 
We adopt the likelihood function of \equref{eq:likelihood} in both cases.

We now discuss a few technical details relevant to our use of the standard template fitting.
First, the likelihood function does not include theoretical uncertainties ($\mat{\Sigma}^*_\mat{F}=0$) since those templates do not have uncertainties. 
Second, we marginalize over $\ell$ with a flat prior, as commonly done with such \photoz methods to focus on color information.
Third, for the 8 templates of the CWW library we use priors of the form $p(z,t_i)= p(z|t_i)p(t_i) = ({a_i}/{b_i} ) z \exp(-{z^2}/{2 b_i} ) $ for $i=1, \cdots, 8$. 
The coefficients are calibrated on the training set following \citep{Coleman:1980ej, Benitez:1998br}, and we find
$\{a_i\} = \{ 0.23, 0.26, 0.32, 0.065, 0.016, 0.067, 0.021, 0.022 \}$ and $\{b_i\} = \{ 0.35, 0.36, 0.37, 0.51, 1.6, 0.38, 0.85, 1.3\}$, for the templates ordered from redder to bluer, \ie spiral and elliptical to irregular galaxies (in this specific order: El\_B2004a, Sbc\_B2004a, Scd\_B2004a, SB3\_B2004a, SB2\_B2004a, Im\_B2004a, ssp\_25Myr\_z008, ssp\_5Myr\_z008).

We now describe more specific details of the novel method and how it is applied to the COSMOS/G10 data.
In the SED model of \equref{eq:sedmodel}, we do not consider a linear mixture of templates, but a single best-fit template per training galaxy instead.
It is obtained by performing standard template fitting and fixing $\alpha_t=0$ except for the best template found. 
We also fix $\ell$ to the best-fit value found with this template.
This speeds up the computation of the overall method, facilitates the interpretation of the results, and alleviates the need to specify priors for the $\alpha_t$ coefficients.

For the redshift-type priors $p(z | t_i) \ p(t_i)$ we adopt a flat prior ($ p(t_i) = $ constant) and a Gaussian prior around the spectroscopic redshift, $p(z | t_i) = \mathcal{N}(z-z_i; \sigma^2_z)$.
The rationale is to only allow training galaxies to contribute to the redshift posterior distribution of target galaxies in a limited redshift range (\ie not all types can affect all redshifts).
The (hyper)parameters of the method are: the parameters of the Gaussian Process (see \appref{sec:rbfgp}), the width of the luminosity and redshift prior $\sigma_\ell$ and $\sigma_z$,  and an extra fractional error which is added in quadrature to all flux errors to compensate for their underestimation if necessary.
We set the later to $1\%$ for the SDSS bands, since those fluxes are known to be reliable at the $1\%$ level.

We perform a rough grid search to find parameters that yield good results on the target.
We use
$\sigma_z = 0.5$, 
$\sigma_\ell = 0.5$,
$\alpha_C= 1000$,
$V_C= 0.5$,
$\alpha_L= 100$,
$V_L= 0.5$,
$\{\mu_i\} = \{ 6500, 5002, 3732 \}$,
$\{\sigma_i\} = \{ 20, 20, 20 \}$.
We expect that the results presented below could be significantly improved by performing a more detailed optimization of the parameters and the use of more complicated priors, for example in terms of redshift and luminosity.
Here we only aim at showing how constructing templates directly from the training set with weak assumptions about galaxy SEDs is straightforward and can improve upon standard template fitting, which relies on templates constructed from low redshift data or complicated physical models.

\begin{figure}
\includegraphics[width=12cm]{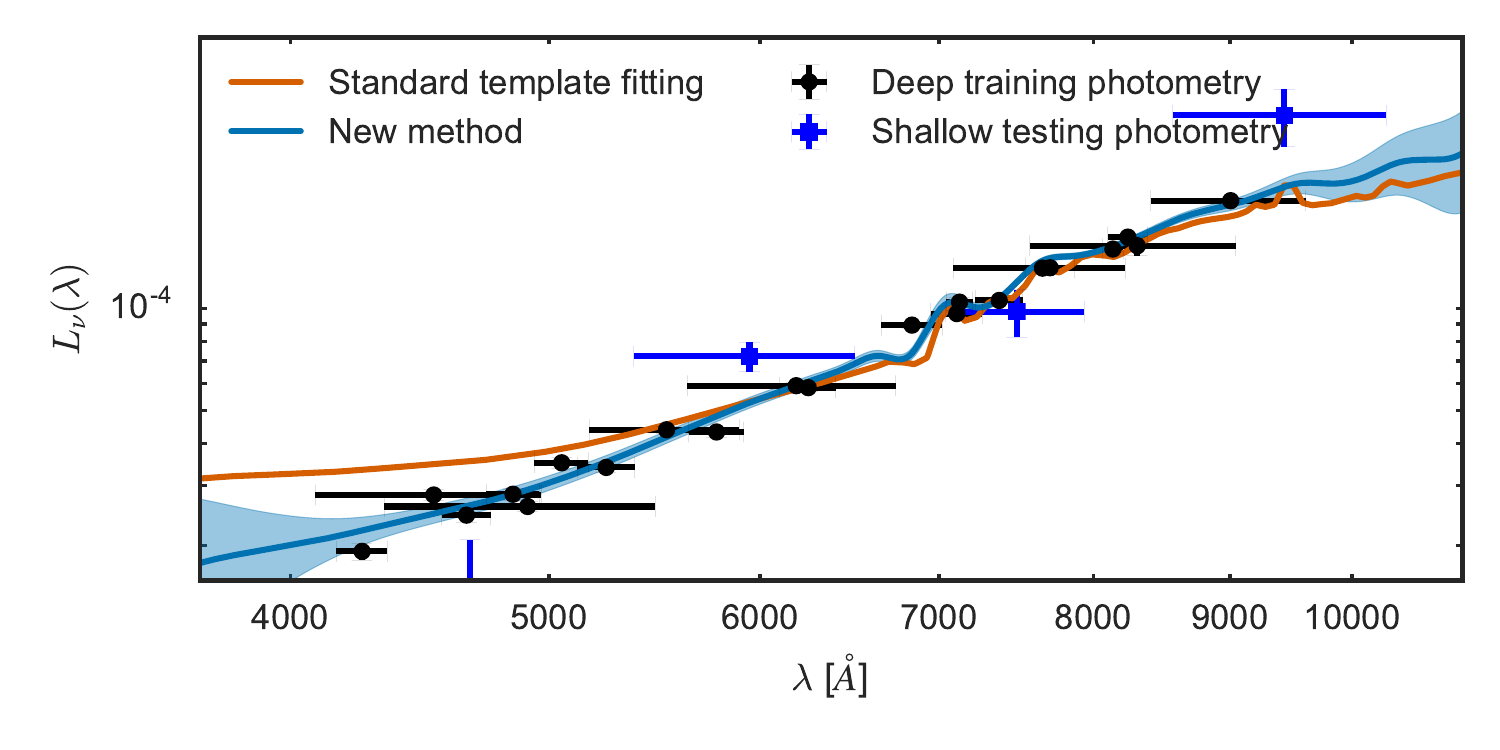}
\caption{SED model constructed for one training galaxy using the deep SUBARU+HST photometric fluxes (black circles). We also show the shallower SDSS fluxes (blue squares), which are not used for constraining the SED. The red line is the best-fit template from our SED library, which would be used in a standard template fitting approach. The blue bands show the 1-sigma errors of space of SEDs constructed with Gaussian Process residuals. Note that in our method this fit in wavelength-SED space is never explicitly performed; instead, our Gaussian Process realizes the same fit but directly in flux--redshift space (thus, the SED model is `latent'), as shown in \figref{fig:traininggalaxy_fluxredshiftmodel}.}
\label{fig:traininggalaxy_fnulambda}
\end{figure}

\begin{figure}
\includegraphics[width=18cm, trim = 2cm 0cm 2cm 0cm, clip]{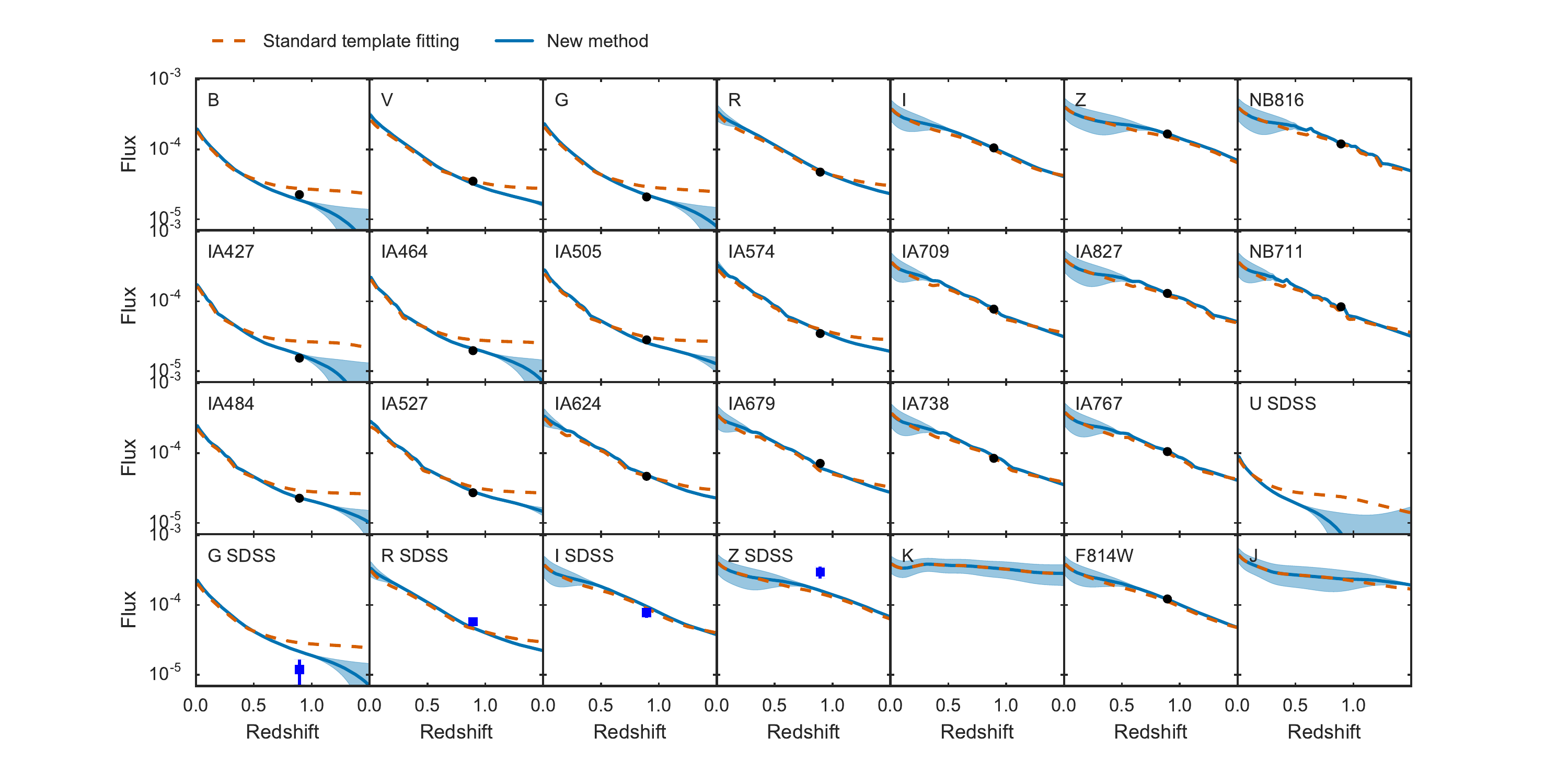}
\caption{Flux--redshift model constructed for the training galaxy of \figref{fig:training_vs_target} obtained by applying our Gaussian Process to SUBARU bands (black dots). The predictions for SDSS bands (blue squares) are also shown. Those flux--redshift envelopes could be equivalently obtained by redshifting and integrating the SEDs shown in \figref{fig:training_vs_target}. But this is made computational tractable by utilizing the Gaussian Process, which implicitly performs the redshifting and projection onto band-passes with a latent SEDs.}
\label{fig:traininggalaxy_fluxredshiftmodel}
\end{figure} 

\begin{figure}
\includegraphics[width=16cm, trim = 0cm 0.6cm 0cm 0cm, clip]{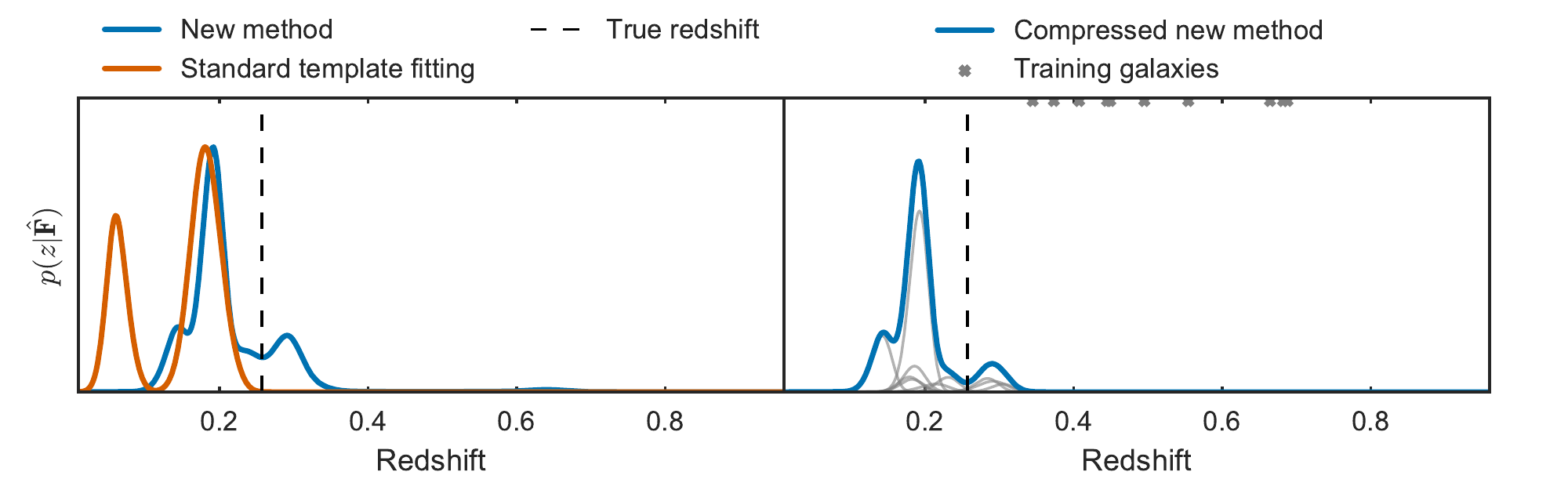}
\includegraphics[width=16cm, trim = 0cm 0.0cm 0cm 1.2cm, clip]{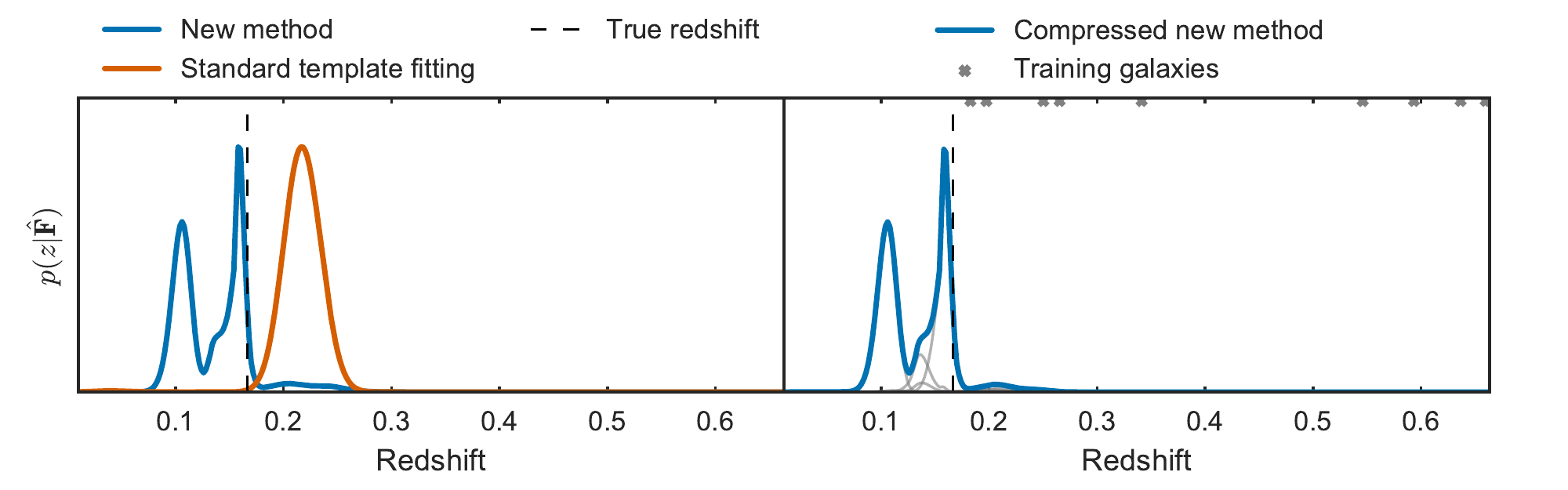}
\caption{Redshift posterior probability distribution functions (PDF) of a few target galaxies, using standard template fitting and the new method. 
The right panels show the ``compressed" PDF obtained with the new method using only the 10 training galaxies contributing the most of the PDFs of the left panels.
The redshifts of these training galaxies, from which we created flux--redshift tracks as in \figref{fig:traininggalaxy_fluxredshiftmodel}, are highlighted with grey crosses, and their PDF contributions with light grey curves. }
\label{fig:datapdfs}
\end{figure}

%%%%%%%%%%%%%%%%%%%%%%%%%%%%%%%%%%%%%%
\subsection{Results}

Before we analyze the results on the full COSMOS/G10 sample, we focus on one training galaxy and examine more closely the model considered and fitted by our method.
As mathematically proved in \secref{sec:gppres}, our Gaussian Process acting in flux--redshift space is equivalent to fitting the SED model of \equref{eq:sedmodel} to noisy photometric fluxes at the spectroscopic redshift.
Let us explicitly perform this fit in wavelength space as an illustration.
We randomly pick a training galaxy and fit the deep 21 bands with the model of \equref{eq:sedmodel}, \ie with the best-fit template from the CWW library, and additive residuals following a Gaussian Process described in \secref{sec:rbfgp}.
Both components are discretized on a fine grid and integrated in the photometric filters, yielding model fluxes which can be compared to observations via a multivariate Gaussian likelihood as described above. 
We draw samples from the posterior distribution on the parameters of this model (the amplitude of the template and discretized residuals) using the \textit{emcee} code \citep{emcee:2013}.
The result of the fit is shown in \figref{fig:traininggalaxy_fnulambda}. 
The initial best-fit template from the CWW library is also shown and is found to be inconsistent with the flux measurements.
As expected, this gets resolved by our inferred solution thanks to the introduction of the Gaussian Process residuals.
But as a result, the model is only well-constrained in the wavelength range covered by the data. 
To predict the fluxes of this galaxy at other redshifts, \ie create model flux--redshift tracks, we could take the samples of the posterior distribution we have just drawn (SEDs), redshift those to a different redshift, and integrate them in the photometric bands.
However, as described above, this process is very computationally intensive, and intractable for a large sample of galaxies. 
Instead, we use the equivalent solution of working directly in flux--redshift space with the Gaussian Process described in \secref{sec:gppres}.
We fit the noisy fluxes of the target galaxy at the spectroscopic redshift, then predict fluxes for other redshifts and bands directly, as detailed in \secref{sec:rbfgp} and also \secref{sec:gppred}.
The result of this process is shown in \figref{fig:traininggalaxy_fluxredshiftmodel} for the same galaxy as \figref{fig:traininggalaxy_fnulambda}.

We see that the flux--redshift relation is well-constrained in a range of redshifts and bands. 
This is not surprising given the broadness and small levels of noise of those 21 bands, which cover a wide range of wavelengths and tightly constrains the shape and amplitude of the SED and thus the flux--redshift relation.
It is only at the highest (lowest) redshift that bluest (reddest) bands are less constrained, as expected from the setup of the problem.

On both figures, we also show the SDSS bands, which are not used in constraining the model but are available for this training galaxy. 
Thus, they are useful for cross-validation.
In this case, we validate that the model trained on the deeper bands successfully predicts the noisier SDSS bands.

\figref{fig:datapdfs} shows two redshift posterior probability distributions (PDFs), calculated with \equref{eq:redshiftposterior} using standard template fitting (8 templates) and our new method (9,923 templates constructed from the training set).
The PDFs look significantly different.
In particular, the new method successfully covers the true redshift of those galaxies, unlike the standard template fitting case.
This is mostly because of the richness of our template space, the propagation of uncertainties from both the SED models and the photometric noise, and also more flexible priors.
The right panels of \figref{fig:datapdfs} shows the PDFs reconstructed from the 10 most significant contributions from training galaxies. 
Since this approximation is essentially indistinguishable from the PDFs in the left panels, we conclude that most of the information in the PDFs comes for a small number of training galaxies \textit{at different redshifts}, also shown in \figref{fig:datapdfs}.
This is a powerful way of compressing the PDFs; we only need to store the indices of the galaxies contributing the most of the PDF. 
The latter can be reconstructed efficiently from the training data and the hyperparameters.

\begin{figure}
\includegraphics[width=18cm]{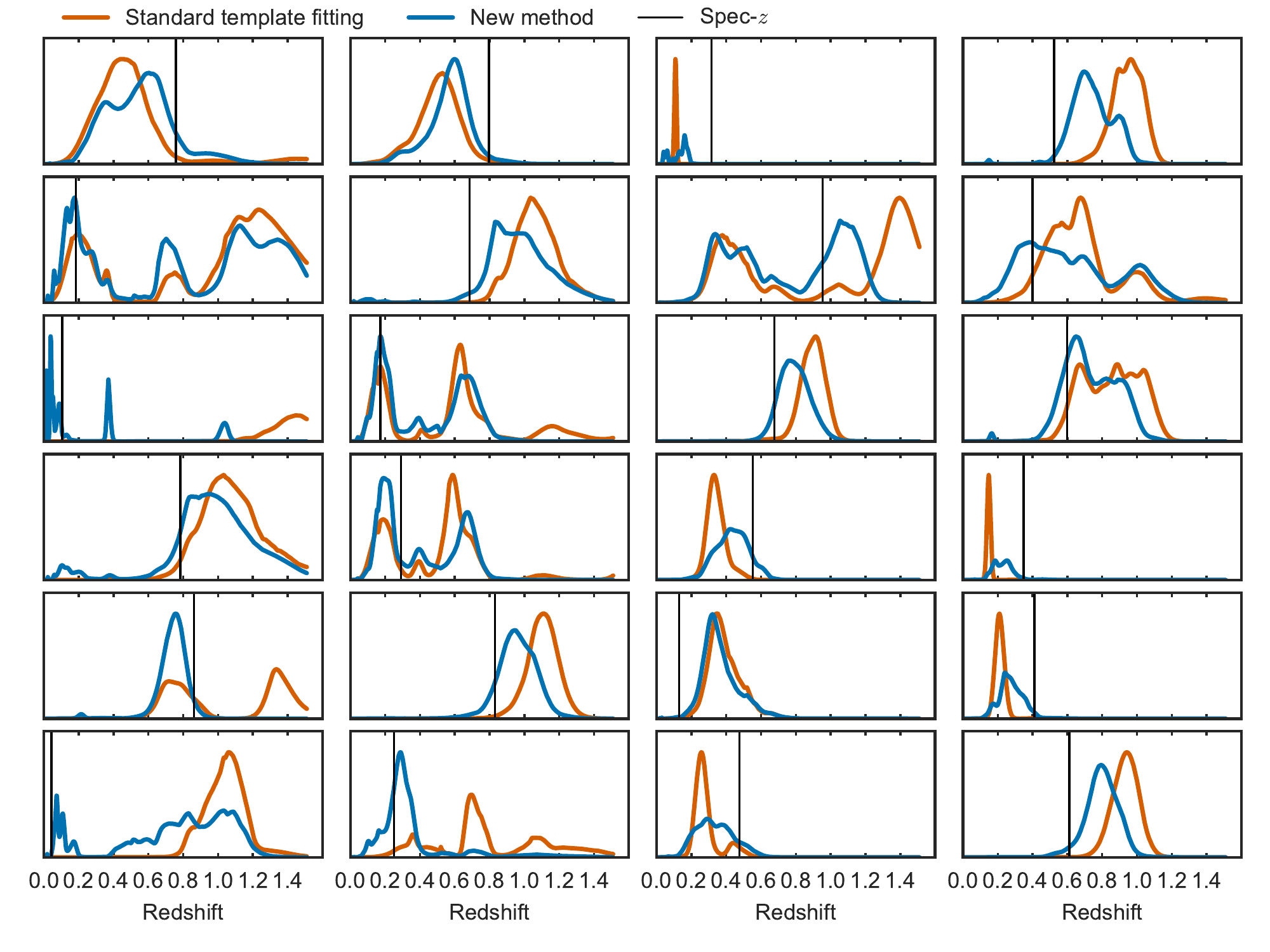}
\caption{Redshift posterior probability distribution functions of a few target galaxies, using standard template fitting and the new method, similar to the left panels of \figref{fig:datapdfs}.}
\label{fig:redshiftpdfs}
\end{figure}

\begin{figure}
\includegraphics[width=16cm, trim = 1.8cm 0cm 3.5cm 0cm, clip]{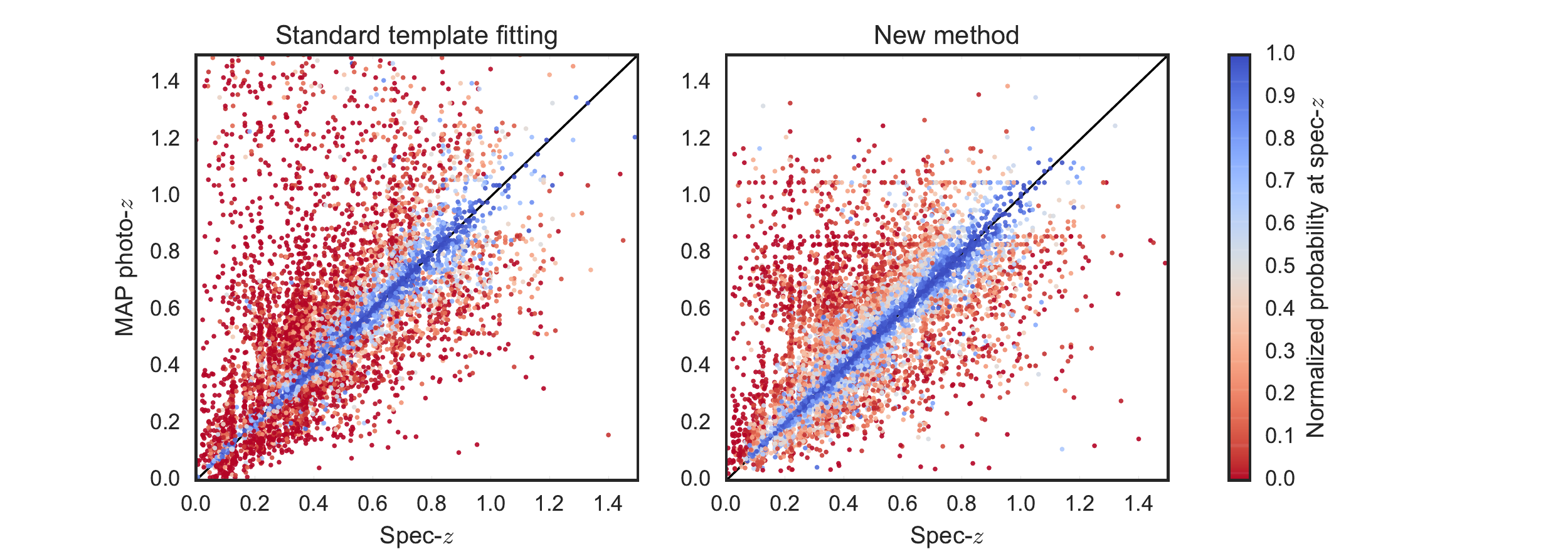}
\caption{Scatter plot of spectroscopic redshifts against the maximum a posterior photometric redshifts (MAP, peak of the posteriors distribution) derived with standard template fitting and the new method. 
The colors show the value of the posterior distribution evaluated at the spectroscopic redshift, with the MAP value normalized to 1.
The new method has a smaller scatter, a smaller fraction of outliers, and produces PDFs that include the true redshifts more frequently.
This is quantified in \figref{fig:cipdfs}.
}
\label{fig:zmean_vs_zspec}
\end{figure}

More redshift PDFs are shown in \figref{fig:redshiftpdfs}, demonstrating that most of them are multimodal and highly structured.
Thus, the mean value of the PDF is not a reliable point estimate.
Instead, we use the maximum a-posteriori (MAP) value, \ie the peak of the PDF.
The distribution of MAP and true redshifts is shown in \figref{fig:zmean_vs_zspec}, for the 8,699 target galaxies and both methods.
The colors correspond to the value of the PDF (normalized to have a MAP value at 1) at the spectroscopic redshifts.
We see that the scatter, number of outliers, and also overall quality of the PDFs is improved with the new method.
The residual outliers are most likely due to systematic biases in the fluxes or their errors. 
Those are problematic for all template fitting methods, and are typically alleviated in machine learning \photoz algorithms.
Our method could be extended to identify and remove inconsistent fluxes.
However, in this paper we only considered the simplest extension and added a constant $1\%$ error in quadrature to all fluxes.

\begin{figure}
\includegraphics[width=18cm]{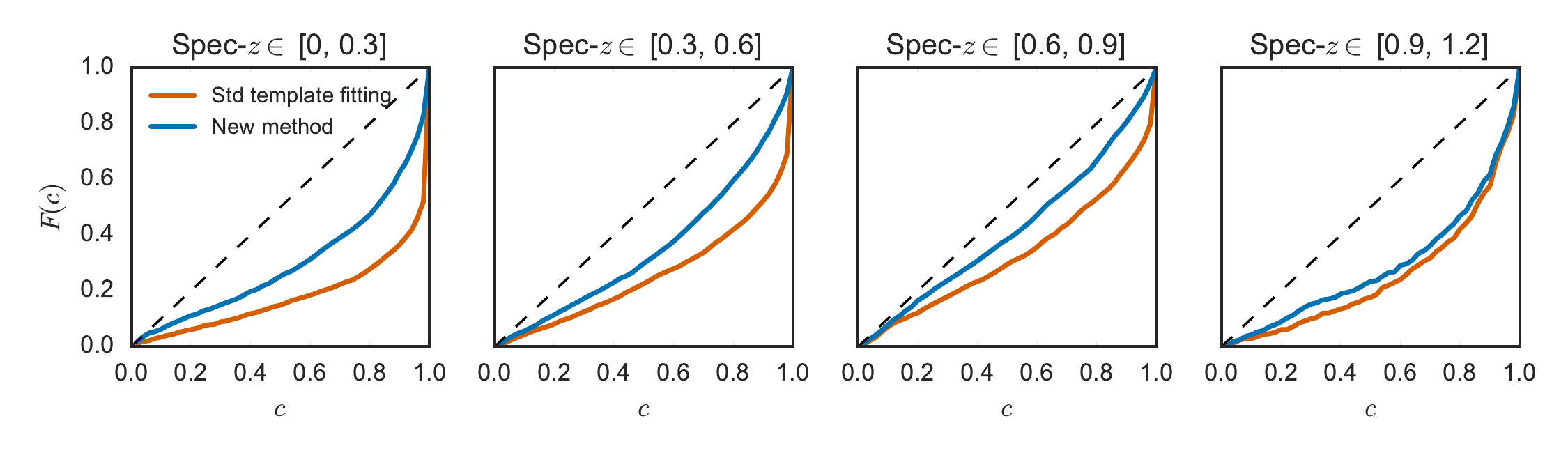}
\caption{Q-Q plots quantifying the quality of the redshift posterior distributions in four redshift bins. 
Results under the diagonal indicate underestimation of the \photoz errors and confidence intervals (see text for details on how we compute $c$).
Note that our method is not optimized and simply consists of performing template fitting with a much larger set of templates directly constructed from the training data.
}
\label{fig:cipdfs}
\end{figure}

To quantify the statistical properties of the PDFs, we divide the target set into four redshift bins (using the spectroscopic redshifts) and evaluate the average quality of the PDF confidence intervals in each redshift bin.
For each galaxy, we find the value of the redshift posterior distribution at the true redshift, and we then integrate the posterior distribution where it is greater than this value.
In other words, for each target galaxy we compute
\equ{
	c = \int_{\mathcal{Z}} \d z \ p(z | \hat{\mat{F}}) 
}
with $\mathcal{Z} = \{ z \ : \ p(z | \hat{\mat{F}})  \ge p(z=z_{\mathrm{spec}} | \hat{\mat{F}}) \}$.
The $c$'s should be uniformly distributed in $[0, 1]$, indicating that the confidence intervals arising from the redshift posterior are statistically correct. 
The cumulative distribution $F(c)$ should follow the $F(c)=c$ diagonal; this is known as a Q-Q plot.  
Q-Q plots for the four redshift bins are shown in \figref{fig:cipdfs}.
We recover a well-known result \citep[\eg][]{Wittman:2016}: standard template fitting significantly underestimate \photoz errors.
This is typically resulted by artificially inflate the PDF, which cannot address the fundamental limitations of the method and reduce extreme outliers due to type degeneracies not captured with the template library.
In our method, we can optimize physically meaningful parameters to improve the confidence intervals.
Even though we have not optimized the parameters in this example, we see that the confidence intervals are more robust.

%%%%%%%%%%%%%%%%%%%%%%%%%%%%%%%%%%%%%%
\section{Conclusion}\label{sec:concl}

Inaccurate photometric redshift estimation is the dominant source of statistical and systematic errors in modern photometric galaxy survey analyses.
This is partly because current \photoz algorithms are unable to model the flux--redshift relation in a flexible, data-driven fashion (\eg via machine learning) while simultaneously exploiting some of our knowledge about the underlying physics (as in template fitting methods).
On the first hand, machine learning methods must be trained on homogeneous, representative data set with the same band-passes, noise and redshift distributions as the target survey of interest.
Thus, they cannot fully take advantage of the wealth of diverse, deep spectroscopic or many-band photometric surveys in existence.
In addition, they do not constrain the flux--redshift relation to correspond to SEDs being redshifted and projected into photometric band-passes. 
On the other hand, libraries of SED templates---which strictly satisfy this constraint---are usually small and constructed from low-redshift data. 
Furthermore, they require significant fine-tuning (of, \eg priors and mapping onto fluxes) to yield robust results for modern data sets. 

To resolve those issues, we presented a new method combining the advantages of both approaches and capable of using heterogeneous training data.
Our method constructs an ensemble of flux--redshift tracks compatible with the noisy fluxes and spectroscopic redshifts of those training galaxies, with the tracks being constrained to correspond to latent SEDs observed through known photometric band-passes.
Thus, it self-consistently accounts for heterogeneity of the noise or redshift distributions, and also missing bands, thanks to the modeling of photometric band-passes and latent SEDs. 
It is made computationally efficient via a Gaussian process for fitting each training galaxy with an ensemble of flux--redshift models and performing a fast comparison with target galaxies. 
This approach is similar to the K-correct model \citep{Blanton:2007}, but constructed from deeper data, including model uncertainties, and does not require explicit construction of the set of SEDs.
The training data can consist of a combination of spectroscopic and many-band photometric data, with arbitrary distributions of flux noise, galaxy types, and redshifts.
Therefore, our method is capable of fully exploiting the setup in which ongoing galaxy surveys operate: the typical training data is highly heterogeneous and extracted from small regions of the survey where more photometric bands and spectroscopic redshifts are available, often from external data. 

Our method is also agnostic to the uniformity of the training set.
We demonstrated this feature on the G10/COSMOS data by training on 9,923 galaxies with 21 deep photometric bands and inferring redshifts for 8,699 galaxies with only 5 noisier SDSS bands available. 
The resulting redshift posterior distributions are statistically robust, as quantified by maximum a-posteriori values and Q-Q plots of the confidence intervals. 
In theory, this method should outperform both template fitting and machine learning methods since it has the advantages of both, while being more flexible and capable of exploiting more data. 
However, our tests were on a relatively small data set with modest statistical power, and we have not pushed the flexibility of the method, especially in terms of optimization of the hyperparameters.
Yet, it is remarkable that it already provides a significant improvement compared to standard template fitting, given that we simply created a large number of flux--redshift tracks with weak assumptions about galaxy SEDs.
Note that our application to G10/COSMOS cannot be performed easily with machine learning because of the use of different bands and noise distributions.
Next, we will perform more realistic tests on a state of the art data set, including a detailed comparison to optimized template fitting and machine learning methods.
We will also quantify the quality of the resulting redshift distributions inferred from those photometric redshifts, which are critical for cosmological analyses.

The method we presented can be extended in a number of ways (some discussed in \secref{sec:discussion}), and we intend to explore some of those in future work as well.
In particular, one could adopt a more realistic latent SED model, for example including more complicated modeling of the continuum and lines, possibly mapping onto physical parameters such as star formation histories \cite[see \eg][]{Miller:2015:gpqso}..
Similarly, instead of one template per training galaxy, one could encode the existence of a low-dimensional space or sequence of galaxy templates in the method, and learn it from the training data.
Finally, applying the methodology to infer quasar redshifts is straightforward, although other complications may arise, such as time variability and stronger degeneracies in color-redshift space. 
The method can also be used beyond \photoz estimation. 
It can be seen as a fast way to fit the noisy flux measurements of an object with a latent SED model and predict fluxes in other band-passes.
Thus, it could be used to predict missing bands or to run cross-validation and quantify the self-consistency between the fluxes measured for an object with respect to an underlying SED model.
Going even further, it could be used to learn filter responses (or correcting existing descriptions) by imposing that the fluxes of a set of galaxies must be self-consistent, for example.
In addition, the set of flux--redshift tracks can be seen as a rich data-driven model of galaxy fluxes and redshifts. 
By drawing objects from this model and adequate redshift, noise or magnitude distributions, one can generate realistic galaxy fluxes containing statistical and systematic errors from real data.
This could be used to populate galaxy mocks or to perform data augmentation, for example to enrich the training sets used in standard machine learning methods; instead of resorting to significantly up-weighting under-represented (\eg high-redshift) objects, which is typically unstable, one could simply enrich the training set and adjust it to be representative of the target survey. 
We intend to explore these diverse applications in future work.

%%%%%%%%%%%%%%%%%%%%%%%%%%%%%%%%%%%%%%
\section{Acknowledgements}

We thank Hiranya Peiris, Daniel Mortlock, Joshua Speagle, Risa Wechsler, Joe DeRose,
Michael Blanton, Daniel Foreman-Mackey, Will Hartley, Ofer Lahav, Robert Lupton, Alex I. Malz, Peter Melchior, Jeffrey Newman, Michael Strauss, Daniela Huppenkothen, and Teresita Suarez Noguez, for useful conversations.

BL was supported by NASA through the Einstein Postdoctoral Fellowship (award number PF6-170154) and by the Simons Foundation (Junior Fellowship, award number 361058).
DWH was partially supported by the NSF (AST-1517237) and the Moore--Sloan Data Science Environment at NYU.

The G10/COSMOS redshift catalogue, photometric catalogue and cutout tool uses data acquired as part of the Cosmic Evolution Survey (COSMOS) project and spectra from observations made with ESO Telescopes at the La Silla or Paranal Observatories under programme ID 175.A-0839. The G10 cutout tool is hosted and maintained by funding from the International Centre for Radio Astronomy Research (ICRAR) at the University of Western Australia. Full details of the data, observation and catalogues can be found in Davies et al. (2015) and Andrews et al. (2016), or on the G10/COSMOS website: \url{http://cutout.icrar.org/G10/dataRelease.php}

%%%%%%%%%%%%%%%%%%%%%%%%%%%%%%%%%%%%%%
\bibliography{bib}

%%%%%%%%%%%%%%%%%%%%%%%%%%%%%%%%%%%%%%
\appendix

%%%%%%%%%%%%%%%%%%%%%%%%%%%%%%%%%%%%%%%%%
\section{Flux likelihood with model uncertainties and scaling parameters}\label{app:fluxlikelihood}

As presented in \equref{eq:likelihood}, our likelihood function comparing observed and model fluxes reads
\eqn{
	p\bigl(\hat{\mat{F}}| z, z_i, \hat{\mat{F}}_i\bigr) &=& \int \d\ell \ \mathcal{N}\left( \hat{\mat{F}} - \ell\  \mat{F}^*(z); \mat{\Sigma}_{\hat{\mat{F}}} + \ell^2 \mat{\Sigma^*_{{\mat{F}}}}(z)  \right) \  \mathcal{N}(\hat{\ell}; \sigma^2_\ell).
}
\appref{sec:gppred} shows how to compute $ \mat{F}^*(z)$ and $\mat{\Sigma^*_{{\mat{F}}}}(z)$ from $z, z_i, \hat{\mat{F}}_i$.
Notice that they are both scaled by $\ell$, to be marginalized over. 
Unfortunately, for this reason there is no analytical solution for this marginalization, unlike for the case where only the mean is scaled.
However, we find that an iterative approach yields satisfactory results.
In particular, we solve and iterate over the following system of equations, with a starting value $\hat{\ell}^2_\mathrm{MAP}(z)  = 0$,
\eqn{
	\mat{\Sigma}(z)&=& \mat{\Sigma}_{\hat{\mat{F}}} \ +\  \hat{\ell}^2_\mathrm{MAP}(z) \mat{\Sigma^*_{{\mat{F}}}}(z) \\
	F_\mathrm{oo}(z) &=& \hat{\mat{F}}^T \mat{\Sigma}(z)^{-1} \hat{\mat{F}} \ +\  \hat{\ell}^2 / \sigma_\ell^2 \\
	F_\mathrm{tt}(z) &=& \mat{F}^{*T} \mat{\Sigma}(z)^{-1} \mat{F}^* \ +\  1/\sigma_\ell^2\\
	F_\mathrm{to}(z) &=&  \mat{F}^{*T} \mat{\Sigma}(z)^{-1} \hat{\mat{F}} \ +\ \hat{\ell}/\sigma_\ell^2 \\
	\hat{\ell}_\mathrm{MAP}(z) &=& F_\mathrm{to}(z)/F_\mathrm{tt}(z) 	.
} 
It typically converges after one iteration. 
We numerically checked for a range of realistic values of the signal and noise that the approximate solution $\hat{\ell}_\mathrm{MAP}$ is very close to the true maximum a-posteriori estimate of $\ell_\mathrm{MAP}$ which would be obtained numerically. 
We also find that the following Gaussian approximation formula is a satisfying approximation for the $\ell$ marginalization of interest,
\eqn{
	p\bigl(\hat{\mat{F}}| z, z_i, \hat{\mat{F}}_i\bigr)  & \approx & \left( (2\pi)^B F_\mathrm{tt}(z)\ \sigma_\ell^2\ \det\mat{\Sigma}(z) \right)^{-\frac{1}{2}} \exp \left( -\frac{1}{2}F_\mathrm{oo}(z) + \frac{1}{2}\frac{F^2_\mathrm{to}(z)}{F_\mathrm{tt}(z)} \right),
}
with $B$ the number of bands, \ie the size of $\hat{\mat{F}}$.
These approximations are typically very good (better than 1\%) is regimes where $\mat{\Sigma}_{\hat{\mat{F}}}$ dominate over $\mat{\Sigma^*_{{\mat{F}}}}(z)$. 
For all purposes this is sufficient for the likelihood function used in this work. 
The regimes where the model uncertainties dominate the data uncertainties will lead to less constrained flux predictions, which typically contribute less to \photoz estimates. 

%%%%%%%%%%%%%%%%%%%%%%%%%%%%%%%%%%%%%%%%%
\section{Our Gaussian Process flux--redshift kernel}\label{sec:rbfgp}

In this appendix we derive a specific model for our latent SED space and the resulting Gaussian Process kernel in flux space of \equref{eq:fluxgp}.
We model the SED residuals of \equref{eq:sedmodel} as
\equ{
	R_\nu(\lambda) = C(\lambda) + \sum_{l} A(\lambda) \mathcal{N}(\lambda-\lambda_l; \sigma^2_l),
}
where $C$ represent the continuum and $A$ the amplitude of Gaussian emission or absorption lines of fixed location and size $ \lambda_l, \sigma_l$ with $l = 1, \ldots, {\rm N}_{\rm lines}$. 
Both $C$ and $A$ are modeled as zero-mean Gaussian Processes with factorized kernels
\eqn{
	C(\lambda) \sim \GP\left(0, k^C(\lambda, \lambda')\right)\\
	A(\lambda) \sim \GP\left(0, k^A(\lambda, \lambda')\right)
}
If one further assumes that $C$ and $A$ are uncorrelated, then 
\eqn{
	R_\nu(\lambda) \sim \GP\left(0, k^R(\lambda, \lambda')\right)
}
with 
\eqn{
	k^R(\lambda, \lambda') = k^C(\lambda, \lambda') + k^A(\lambda, \lambda')\sum_{l} \mathcal{N}(\lambda-\lambda_l;\sigma^2_l) \ \mathcal{N}(\lambda'-\lambda_l; \sigma^2_l)
	}
assuming that the various lines $\mathcal{N}(\lambda, \lambda_l, \sigma^2_l)$ don't overlap significantly so that cross terms can be neglected. 
Those assumptions could be relaxed, but they greatly simplify the calculations and have a small effect on the resulting fluxes in the case of broad photometric bands.

We now consider the specific case of all kernels being Gaussian (which is sometimes called a Radial Basis Function in the Gaussian Process litterature),
\eqn{
	k^C(\lambda, \lambda') &=&  V_C\  \mathcal{N}\left(\lambda - \lambda'; {\alpha}^2_C\right) \\
	k^L(\lambda, \lambda') &=&  V_L\ \mathcal{N}\left(\lambda - \lambda'; {\alpha}^2_L\right).
}

We approximate the rescaled filters $V_b(\lambda) = W_b(\lambda)/\lambda$ as a sum of Gaussian distributions,
\eqn{
	V_b(\lambda) = \sum_{i} a_{i}\mathcal{N}(\lambda - \mu_i;\sigma^2_i) \quad\quad  V_{b'}(\lambda) = \sum_{i'} a_{i'}\mathcal{N}(\lambda - \mu_{i'}; \sigma^2_{i'})
}
We drop the secondary $b$ dependency below, but it should be understood that the $i$ and $j$ indices below depend on each band.
The kernel for our flux--redshift Gaussian Process of \equref{eq:fluxgp} is
\eqn{
	k^F(b,b',z,z',\ell,\ell') &=& \left( \frac{ (1+z)(1+z') }{4\pi D(z) D(z') g^\mathrm{AB}} \right)^2 \frac{\ell \ell'}{C_bC_{b'}} \nonumber \\
	  \times \ \sum_{i}\sum_{i'} a_{i} a_{i'} &&\hspace*{-5mm} \left( 2\pi\sigma_{i}\sigma_{i'} V_C\ k^C(z,z',b,b',i,i') + V_L \ \sum_{l}\sum_{l'} k^L(z,z',b,b',i,i',l,l')\right)\label{eq:actualkernel}.
}
The contribution from the continuum is
\eqn{
	k^C(z,z',b,b',i,i') = \frac{{\alpha}_C}{\sigma_{ii'}} \mathcal{N}\left( \mu_{i}(1+z') - \mu_{i'}(1+z); \sigma^2_{ii'}\right),
}	
with
\equ{
	\sigma_{ii'}^2 \ =\ \sigma_{i}^2(1+z')^2 + \sigma_{i'}^2(1+z)^2 + {\alpha}_C^2(1+z)^2(1+z')^2,
}
and the contribution from the lines is
\eqn{
	k^L(z,z',b,b',i,i',l,l') \ = \  \mathcal{N}\left( \mu_{i} - \mu_{l}(1+z); \sigma^2_{i}\right) \ \mathcal{N}\left( \mu_{i'}-\mu_{l'}(1+z'); \sigma^2_{i}\right) \ \mathcal{N}\left( \mu_{l}-  \mu_{l'}; {\alpha^2_L}\right).
}	
Our kernel is is parametrized with $(V_L,  \alpha_L, V_C, \alpha_C)$ controlling the variance and smoothness of the continuum and line residuals. Recall that the filters and the lines are described as Gaussian mixtures, assumed to be known, although they could be set as parameters and also inferred from the data.

%%%%%%%%%%%%%%%%%%%%%%%%%%%%%%%%%%%%%%%%%
\section{Gaussian Process predictions}\label{sec:gppred}

As detailed in the main text, we construct one Gaussian Process per training galaxy, and using noisy flux measurements and the spectroscopic redshift we make model predictions for other fluxes at different redshifts, via \equref{eq:likelihood}. 
To simplify the notation we drop the subscript $i$ from the main text; otherwise it should be added to all the variables below.
Specifically, the training galaxy has (spectroscopic) redshift $z$ and noisy photometric fluxes $\hat{\mat{F}} = (\cdots, \hat{F}_{b_j}, \cdots)$ with $j=1, \dots, B$ with $B$ the number of observed bands and $b_j$ the labels of the bands. 
The covariance of these measured fluxes is $\mat{\Sigma}_{\hat{\mat{F}}}$, and $\hat{\ell}$ is the estimated absolute luminosity (see main text for details).

We will predict noiseless fluxes for the same training galaxy at a different redshift $z^*$ and at absolute luminosity $\ell^*$. 
We will use ${\mat{F}}^* = (\cdots, {F}^*_{b_k}, \cdots)$ with $k=1, \dots, B^*$ with $B^*$ the number of predicted bands. 
One of the advantages of our approach is that it does not require the sets of bands $\{ b_j \}_{j=1, \dots, B}$ and $\{ b_k \}_{k=1, \dots, B^*}$ to match; we can make predictions for arbitrary bands using any set of measured bands as long as the filter responses are perfectly known. Note that this could be relaxed by inferring corrections to the filter responses directly from the data within our framework.

We will follow the language of Gaussian Processes and use three input dimensions (luminosity $\ell$, band $b$, and redshift $z$) and one output dimension (photometric flux). Note that the second input dimension takes discrete values in the set of possible bands under consideration (the union of all possible bands for all training and target galaxies).

Thus, the Gaussian Process is trained to fit the vector $\hat{\mat{F}}$ of size $B \times 1$ given the input matrix $\mat{X}$ of size $B \times 3$, with the $j$th row being the 3-element vector
\eqn{
	\mat{X}_j &=& (b_j,\ z,\ \hat{\ell}).
}
Similarly, we will make prediction for the vector ${\mat{F}}^*$ of size $B^*\times 1$ given the input matrix $\mat{X^*}$ of size $B^* \times 3$, with the $k$th row being the 3-element vector
\eqn{
	\mat{X}^*_k &=& (b_k^*, \ z^*, \ {\ell}^*) .
}
The prior on the noiseless model fluxes $\mat{F}$ (before observing the data $\hat{\mat{F}}$) reads
\eqn{
	p( \mat{F} | \mat{X} ) =  \mathcal{N}(\mu^F( \mat{X}) ; k^F(\mat{X},\mat{X})) ,
}
and the posterior distribution on $\hat{\mat{F}}^*$ given measured $\hat{\mat{F}}$  is
\eqn{
	p( \hat{\mat{F}}^* | \mat{X}^*, \hat{\mat{F}}, \mat{X} ) =  \mathcal{N}({\mat{F}}^*; \mat{\Sigma}_{\mat{F}^*}) 
}
with mean and covariance
\eqn{
	\mat{F}^* &=& \mu^F(\mat{X}^*) \ + \ k^F(\mat{X}^*, \mat{X}) \ \bigl[ k^F(\mat{X},\mat{X})+ \mat{\Sigma}_{\hat{\mat{F}}} \bigr]^{-1} \bigl( \hat{\mat{F}} - \mu^F( \mat{X}) \bigr) \\
	\mat{\Sigma_{{\mat{F}}}^*}&=&  k^F(\mat{X}^*, \mat{X}^*) \ -\ k^F(\mat{X}^*, \mat{X}) \ \bigl[k^F(\mat{X}, \mat{X}) + \mat{\Sigma}_{\hat{\mat{F}}} \bigr]^{-1} k^F(\mat{X}, \mat{X}^*).
}
In these equations, $\mu^F(\mat{X})$ and $\mu^F(\mat{X}^*)$ are $B \times 1$ and  $B^* \times 1$ vectors denoting the mean function of \equref{eq:gpmeanfct} evaluated at the inputs $\mat{X}$ and $\mat{X}^*$, respectively. 
Similarly, $k^F(\mat{X}, \mat{X})$, $k^F(\mat{X}^*, \mat{X})$ and $k^F(\mat{X}^*, \mat{X}^*)$ denote the kernel of \equrefs{eq:gpgenkernel}{eq:actualkernel} evaluated at the inputs $\mat{X}$ and $\mat{X}^*$, and have size $B \times B$, $B^* \times B$, and $B^* \times B^*$, respectively.

%% %%%%%%%%%%%%%%%%%%%%%%%%%%%%%%%%%%%%
\end{document}